\documentclass{myappolb}
\usepackage{epsfig}
\newcommand{\GeV}{\rm GeV}

\newcommand{\xg}{$x_\gamma$\,}
\newcommand{\Et}{$\overline{E}_t^2$\,}
\newcommand{\qq}{$Q^2$\,}
\newcommand{\EtEQ}{\overline{E}_t^2}
\newcommand{\gammaT}{$\gamma^*_T$\ }
\newcommand{\gammaL}{$\gamma^*_L$\ }
\begin{document}
\title{Dijet Production at Low Q$^2$
\thanks{Presented at DIS 2002 conference.}%
}
\author{Kamil Sedl\'ak
\address{Center for Particle Physics,
Institute of Physics, AS CR, \\
Na Slovance 2, 182 21 Praha 8, Czech Republic
\\E-mail: ksedlak@fzu.cz\\{\rm On 
behalf of the H1 Collaboration}}
}
\maketitle
\begin{abstract}
Triple differential dijet cross-sections in $e^\pm p$ interactions measured with
the H1 detector at HERA are presented. The data are compared to Monte
Carlo simulations based on the DGLAP and CCFM parton evolution sche\-mes.
Effects of longitudinally polarized virtual photons are investigated.
\end{abstract}
  
\section{Dijet Production at HERA}
\label{uvod}
The production of dijet events at HERA is dominated by processes
in which a virtual photon, coupling to 
the electron, interacts with a parton in the proton.
In the region of photon virtuality $Q^2 \gg \Lambda_{\mathrm QCD}^2$, hard collisions
   of the photons do not necessitate the introduction of the concept
   of the resolved photon (as for the real photon) and the process can in
   principle be described by the direct photon contribution alone.

The analysis presented here explores the region 
   $\Lambda_{\mathrm QCD}^2 \ll Q^2$ \raisebox{-0.8ex}{$\stackrel{\mathrm{<}}{\sim}$} 
$E_t^2$, where different theoretical approaches can
   be used to take into account higher order corrections
   -- either next-to-leading order (NLO) calculations,
   $k_t$ unordered initial QCD cascades, or 
   additional interactions with resolved photons. 
   Comparisons with NLO predictions are
   likely to become possible in the future.  In more detail, 
   the present measurements are compared with the following models:

\vspace{0.8em}
{\bf a) LO direct and resolved interactions} based on the DGLAP 
     evolution equations and parton showers.
     The effects of transversally ($\gamma^*_T$) and also
     longitudinally ($\gamma^*_L$) polarized resolved photon 
     interactions are studied~\cite{Sedlak,chyla,Chyla:2000hp}.
     The cross section for longitudinal photons vanishes 
     for $Q^2 = 0$ due to gauge invariance. 
     On the other hand, the concept of a resolved photon 
     breaks down for $Q^2 > E_t^2$. Therefore
     the most promising region in which to search for the $\gamma^*_L$ 
     resolved processes is
     $\Lambda_{\mathrm QCD}^2 < Q^2 \ll E_t^2$, 
     which is often the case in the present analysis.
     The main difference between $\gamma^*_L$
     and $\gamma^*_T$ induced interactions
     arises from the $y$ dependence of the respective fluxes:
     \begin{eqnarray}
       \label{rovnice1}
       f_{\gamma^T/e}(y,Q^2) & 
       = & \frac{\alpha}{2\pi} \left[  \frac{2(1-y)+y^2}{y}
         \frac{1}{Q^2} - \frac{2m_e^2 y}{Q^4} \right] \\
       \label{rovnice2}
       f_{\gamma^L/e}(y,Q^2) & 
       = & \frac{\alpha}{2\pi} \left[  \frac{2(1-y)}{y}
         \frac{1}{Q^2}  \right]
     \end{eqnarray}
     While for $y\rightarrow 0$,
     both transverse and longitudinal fluxes are approximately same,
     the longitudinal flux vanishes for $y\rightarrow 1$.
     Also the dependence of the 
     point-like~\footnote{The perturbatively non-calculable 
     hadron-like part of the photon PDF 
     becomes negligible in our kinematical region with respect
     to the point-like one, 
     as has been demonstrated in~\cite{Chyla:1999pw}.}
     (i.e. perturbatively calculable)
     parts of the photon parton distribution functions (PDF) 
     on \qq and $E_t^2$
     differs -- while the $\gamma^*_T$ PDF 
     are proportional to $\ln (E_t^2/Q^2)$, 
     the $\gamma^*_L$ PDF do not, in the first approximation, 
     depend on either $E_t^2$ or \qq~\cite{chyla}.

\vspace{0.8em}
{\bf b) $\mathbf k_t$ unordered initial QCD cascades} accompanying the hard 
     process are present for example in BFKL or CCFM evolution. These evolution 
     schemes can lead to final 
     states in which the partons with the
     largest $k_t$ may come from the cascade, and not, as in DGLAP
     evolution, from the hard subprocess. Such 
     events may have a similar topology to that for the resolved 
     interactions in the DGLAP approximation.
     This possibility is investigated using the CASCADE~1.0 
     generator~\cite{Cascade,Cascade2,Unintegr}
     based on the CCFM evolution equations.

%
%
\section{Measurement of Dijet Cross-Section}

The measurement was done with 16.3 pb$^{-1}$ of data collected in 1999, when
the electron-proton center-of-mass energy $\sqrt{s}$ reached 318~\GeV.
The ana\-lysis was performed in $\gamma^*$-proton center-of-mass system
and jets were found using the $k_t$ longitudinally invariant jet algorithm.
The phase space is defined by the photon virtuality:
 $2~{\GeV}^2 < Q^2 < 80~{\GeV}^2$, the electron inelasticity: $0.1 < y < 0.85$,
the transverse energy of two leading jets: $E_t^{jet\,1,2} > 5~\GeV$, 
$\overline{E}_t=(E_t^{jet\,1}+E_t^{jet\,2})/2 > 6~\GeV$ and
the pseudorapidity of the two leading jets:
$-2.5 < \eta^{jet\,1,2} < 0$.

The measured data are corrected for detector effects using  
the Bayesian unfolding method.  The largest source of
systematic
errors arises from the model dependence of the detector 
correction, and from
the main calorimeter calibration uncertainty.

\section{Results and Discussion}

\begin{figure} \centering 
  \epsfig{file=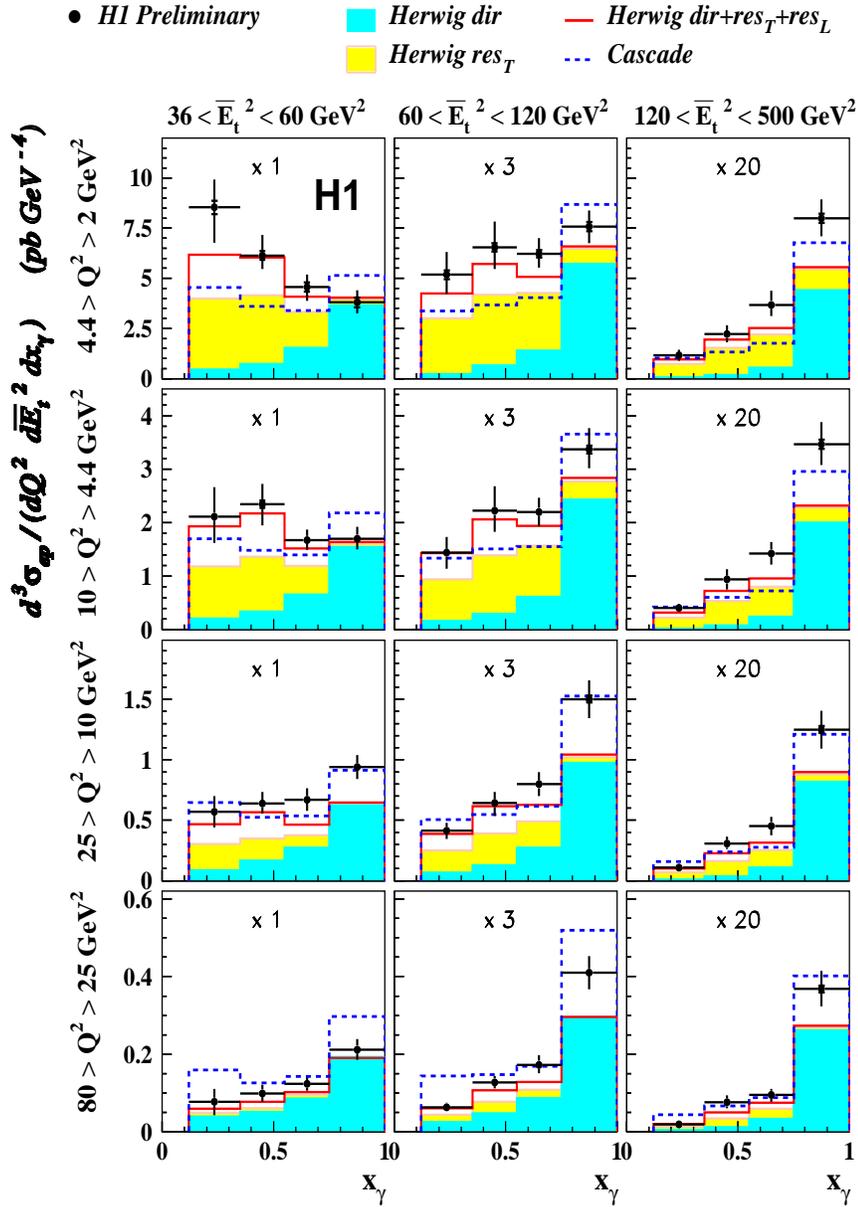,width=126mm,height=174mm}
  \caption{Triple differential dijet cross-section 
    $d^3\sigma_{ep}/dQ^2 d\overline{E}_t^2 dx_\gamma$ for the H1 data
    depicted by points is compared to predictions of the HERWIG and
    CASCADE MC programs.}
  \label{H1a}
\end{figure}

The corrected triple-differential dijet cross-section measured
as a function of \qq, \Et and \xg is shown in
Fig.~\ref{H1a}.
A prediction of HERWIG~\cite{Herwig} with the SaS1D parameterization of 
the \gammaT PDF, as well as the pure direct contribution
is compared to the data.

In general, HERWIG~5.9 and RAPGAP~2.8~\cite{Rapgap} tend to underestimate the 
measured cross-section.
The decrease of the resolved contribution at high \Et is
of kinematic origin, due to the limited phase space
at low \xg.  The direct contributions almost describe the data
in the highest \qq bin, while a clear need for resolved processes is
observed for $Q^2 \ll \EtEQ$.

In the
highest \qq range ($25 < Q^2 < 80 ~\GeV^2$) and $x_\gamma<0.75$, 
the HERWIG direct contribution almost describes the data 
in the lowest \Et bin, but is significantly below it 
in the highest \Et bin.  
This indicates that the relevance of the resolved photon contribution 
is governed by the ratio $\overline{E}_t^2 /Q^2$, 
rather than by $Q^2$ itself.

Standard HERWIG with direct and \gammaT resolved
contributions underestimates the data. The description 
is improved by adding \gammaL  resolved photon 
interactions,
which is done using a slightly
modified version of HERWIG with the longitudinal photon 
flux according to eq.~(\ref{rovnice2}) and 
a recent $\gamma^*_L$ PDF parameterization~\cite{Chyla:2000hp}.
As demonstrated in Fig.~\ref{H1a}, the \gammaL resolved
contribution is significant, and brings  HERWIG
closer to the measurement.

\begin{figure} \centering 
  \epsfig{file=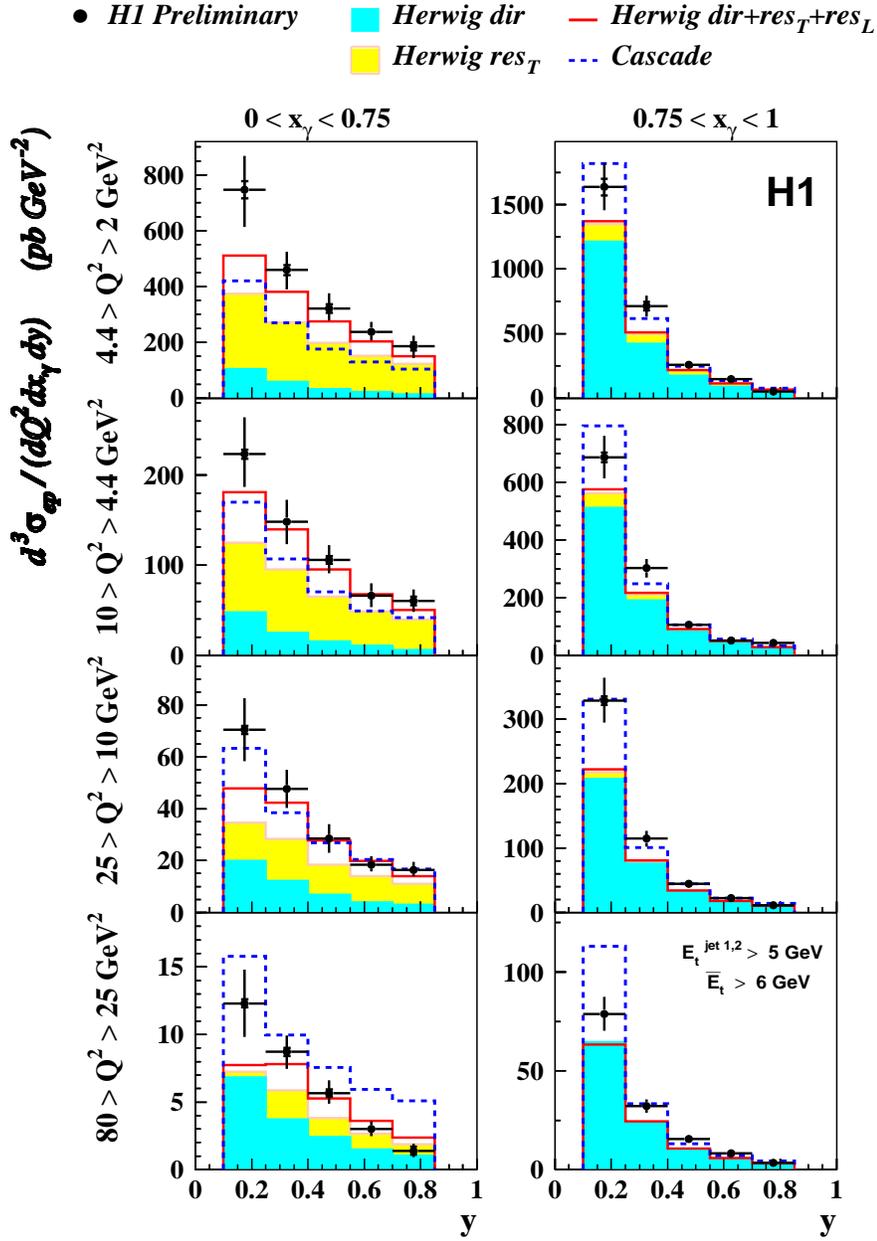,width=126mm,height=180mm}
  \caption{Triple differential dijet cross-section
    $d^3\sigma_{ep}/dQ^2 dx_\gamma dy$.}
  \label{H1b}
\end{figure}

On the other hand, a simple enhancement of the PDF of the $\gamma^*_T$ 
in the resolved contribution could lead to a similar prediction
as the introduction of resolved $\gamma^*_L$. 
To distinguish between a non-optimal choice of \gammaT PDF and
the need for resolved \gammaL,
the dijet cross-section has also been studied
as a function of \qq, \xg\, and $y$, which is shown in Fig.~\ref{H1b}.
HERWIG is below the data. The discrepancy becomes smaller 
if the resolved \gammaL is added. 
According to eq.~(\ref{rovnice1}-\ref{rovnice2}),
the slope of inelasticity $y$ of the HERWIG prediction 
in the region of $x_\gamma<0.75$, depends significantly on 
whether $\gamma^*_L$ processes are included or not. 
Unlike a pure enhancement of $\gamma^*_T$ PDF, which would not
change the slope of the $y$ distribution, addition of $\gamma^*_L$ brings
the $y$ dependence of HERWIG much closer to the measurement.

As motivated in Section~\ref{uvod},
the measured cross-sections  are also compared 
to a prediction of the CASCADE MC program based on the CCFM evolution
scheme. 
This theoretical approach does not involve 
the concept of virtual photon structure and employs
much fewer parameters for tuning than the usual DGLAP-based MC
programs. 
CASCADE describes the data reasonably
but not perfectly.
In particular, the \qq dependence at low \xg is poorly described.

As indicated by Fig.~\ref{H1b}, the $y$ dependence of the
dijet cross-section is better described by CASCADE than by 
HERWIG without the $\gamma^*_L$ resolved process, since
photon polarization states are correctly treated in CASCADE
for all virtualities (only direct photon interactions are considered).
\section{Conclusions}

The importance of \gammaT resolved
photon interactions within the DGLAP evolution scheme
at leading order is clearly demonstrated in the region where
$\EtEQ > Q^2$, even at rather high $Q^2$.
Additional \gammaL resolved photon contributions
further improve the agreement of HERWIG with the
measured data.

  Exploring the CCFM approach, the MC program CASCADE 
does not reproduce the data perfectly, the main discrepancy 
is observed in the \qq de\-pen\-dence at low \xg.
On the other hand, 
the \xg dependence in CASCADE is comparable to the sum of the direct and
resolved contributions in DGLAP-based MC programs, 
showing that non $k_t$ ordered parton cascades
can successfully produce the same observables as resolved virtual photons
in the LO DGLAP evolution scheme.

\section*{Acknowledgments}
I am grateful to J.~Cvach, J.~Ch\'yla, H.~Jung, P.R.~Newman, M.~Ta\v{s}evsk\'y and 
A.~Valk\'arov\'a for many valuable discussions and careful reading
of the manuscript. 

This work has been supported in part by the Ministry of Education 
of the Czech Republic under the project LN00A006.


%
\vspace*{-2cm}

\end{document}